\documentclass[nofootinbib,aps,a4paper,11pt,letterpaper,superscriptaddress,column,showkeys,times,eqsecnum]{revtex4-2}
\usepackage{amsmath}
\usepackage{amsfonts}
\usepackage{graphicx}
\usepackage{color}
\usepackage{amsthm}
\usepackage{amsmath}
\usepackage{braket}
\usepackage{amsmath}
\usepackage{dcolumn}
\usepackage{bm,url}
\usepackage{mathrsfs}
\usepackage[linktocpage]{hyperref}
\usepackage{subfigure}
\usepackage{amsfonts}
\usepackage[left=2.1cm,right=2.1cm,top=2cm,bottom=2cm]{geometry}
\usepackage[usenames,dvipsnames,svgnames]{xcolor}  %% colored text
\usepackage{hyperref}   %% Needs to make hyper reference of any references or citations
\definecolor{oxfordblue}{rgb}{0.0, 0.13, 0.28}
\definecolor{burgundy}{rgb}{0.5, 0.0, 0.13}
\definecolor{darkolivegreen}{rgb}{0.33, 0.42, 0.18}
\definecolor{darkblue}{rgb}{0,0,0.5}
\definecolor{richcarmine}{rgb}{0.84, 0.0, 0.25}
\definecolor{darkblue}{rgb}{0,0,0.5}
\definecolor{bluer}{rgb}{0.00,0.50,0.75}{}
\hypersetup{colorlinks=true, citecolor=red, linkcolor=blue,
urlcolor = magenta, filecolor=magenta}

\begin{document}

\newcommand\be{\begin{equation}}
\newcommand\ee{\end{equation}}
\newcommand\bea{\begin{eqnarray}}
\newcommand\eea{\end{eqnarray}}
\newcommand\bseq{\begin{subequations}} %solo con amsmath
\newcommand\eseq{\end{subequations}}
\newcommand\bcas{\begin{cases}}
\newcommand\ecas{\end{cases}}
\newcommand{\p}{\partial}
\newcommand{\f}{\frac}

\title{Cosmological coupled black holes immersed in dark sector}
\author{Chen-Hao Wu}
\email{chenhao\_wu@nuaa.edu.cn}
\affiliation{College of Physics, Nanjing University of Aeronautics and Astronautics, 
Nanjing,
210016,
China}
\affiliation{Center for the Cross-disciplinary Research of Space Science and Quantum-technologies (CROSS-Q), NUAA, Nanjing, 210016, China}
\author{Yue Chu}
\email{ychu77@nuaa.edu.cn}
\affiliation{College of Physics, Nanjing University of Aeronautics and Astronautics, 
Nanjing,
210016,
China}
\affiliation{Center for the Cross-disciplinary Research of Space Science and Quantum-technologies (CROSS-Q), NUAA, Nanjing, 210016, China}
{\author{Ya-Peng Hu}
\email{huyp@nuaa.edu.cn}
\affiliation{College of Physics, Nanjing University of Aeronautics and Astronautics, 
Nanjing,
210016,
China}%{College of Physics, Nanjing University of Aeronautics and Astronautics, Nanjing, 210016, China}%Department and Organization
\affiliation{Center for the Cross-disciplinary Research of Space Science and Quantum-technologies (CROSS-Q), NUAA, Nanjing, 210016, China}
\affiliation{MIIT Key Laboratory of Aerospace Information Materials and Physics,  Nanjing University of Aeronautics and Astronautics, Nanjing, 210016, China}

%\date{\today}
%%%%%%%%%%%%%%%%%%%%%%%%%%%%%%%
\begin{abstract}
Motivated by theoretical and observational developments of cosmological coupled black holes, we construct an exact analytical solution for a black hole immersed in an anisotropic dark sector background, adopting the framework established by [Cadoni et al., JCAP 03 (2024) 026]. By generalizing a static seed metric to a dynamical FLRW background, we derive a solution where the black hole mass co-evolves with the cosmic expansion. We then obtain the explicit form of the radius-dependent coupling exponent, revealing that the interaction is governed by the dark halo profile. Considering the ubiquity of the dark halos surrounding supermassive black holes, our model provides a potential realization of cosmological coupling, interpreting the mass growth as the dynamical response of the surrounding dark sector fluid to the Hubble flow, distinct from the method of modifying the black hole's internal equation of state.
	
%\vspace{0.5cm}
%\textbf {Keywords:} 
\end{abstract}

\maketitle
\section{Introduction}
%\label{introduction}
Black holes (BHs) are among the most fundamental predictions of General Relativity (GR). While the existence of BHs has been robustly confirmed by gravitational wave detections \cite{LIGOScientific:2016aoc, LIGOScientific:2017ycc} and the Event Horizon Telescope observations \cite{EventHorizonTelescope:2019dse, EventHorizonTelescope:2019ggy, EventHorizonTelescope:2022wkp}, the standard theoretical descriptions, such as the Schwarzschild and Kerr metrics, treat these objects as isolated systems in an asymptotically flat background. However, our cosmological background is not static, expanding, and is currently dominated by the dark sector, i.e., dark matter and dark energy. This discrepancy between the local static approximation and the global time-dependent background raises a natural question: How do BHs interact with and evolve in an expanding universe?

Historically, the first attempt to embed a BH in a cosmological background was the McVittie solution \cite{McVittie:1933zz}. However, for standard singular BHs in McVittie-type solutions, the BH mass effectively decouples from the cosmic expansion \cite{Cadoni:2023lqe, Gaur:2023hmk}, maintaining a constant ADM mass despite the background evolution. Recently, this theoretical landscape has been challenged by observational evidence. A study by Farrah et al.\cite{Farrah:2023opk} analyzed the growth of supermassive BHs (SMBHs) in elliptical galaxies, revealing that the BH masses increase significantly more than expected from standard accretion or merger channels, and are consistent with a phenomenological law involving scale factor $a$ as \cite{Croker:2021duf}
\begin{equation}
M(a)\propto a^k,\quad-3\leq k\leq3\mathrm{~.}
\end{equation}
It is worth noting that the present observational evidence pointing to this effect is rather controversial \cite{Rodriguez:2023gaa, Andrae:2023wge, Lei:2023mke, Amendola:2023ays, Lacy:2023kbb, Lei:2025qff}. From a theoretical point of view, investigating the dynamical interaction between small and large scales is interesting per se \cite{BenAchour:2025vur}, and such observational findings provide a potential astrophysical motivation for constructing exact GR solutions that exhibit cosmological coupling. If BHs indeed grow with the universe ($k>0$), they cannot be described by simple vacuum solutions or standard dust embeddings. The coupling requires a non-trivial energy-momentum tensor, likely involving anisotropic fluids or exotic matter states, to mediate the interaction between the local compact object and the global Friedmann-Lemaître-Robertson-Walker (FLRW) background \cite{Cadoni:2023lqe, Cadoni:2023lum}.

Meanwhile, considerable astrophysical observations indicate that SMBHs in the giant disk galaxies or elliptical galaxies are surrounded by dark halos \cite{EventHorizonTelescope:2019dse, EventHorizonTelescope:2019ggy, EventHorizonTelescope:2022wkp, Rubin:1980zd, Persic:1995ru, Faber:1979pp, Massey:2010hh, Bertone:2016nfn}. It is physically natural to investigate BHs immersed in a dark Sector environment. Specifically, the simplest case is a spherically symmetric BH immersed in a perfect fluid dark matter (PFDM), combined with interacting dark fluids, which was first provided in Ref.\cite{Li:2012zx}. Considering its simplicity, this solution is widely investigated in different physical contexts \cite{Xu:2017bpz, Haroon:2018ryd, Konoplya:2019sns, Jusufi:2019nrn, Narzilloev:2020qtd, Rizwan:2018rgs, Stuchlik:2019uvf, Hendi:2020zyw, Cao:2021dcq, Xu:2016ylr, Shaymatov:2020wtj, Gao:2023ltr, Liang:2023jrj, Abdusattar:2023xxs, Liu:2024qso, Sadeghi:2024wib, Rodriguez:2024jzw, Narzilloev:2024fsw, Das:2023ess}. Whereas the above studies are all based on a static metric, the cosmological couple effects still remain poorly explored. It is important to emphasize that cosmological coupling need not be restricted solely to dark matter environments. Recent developments \cite{Cadoni:2024jxy} suggest that cosmological coupling might be a more universal property of local gravitational systems (see also previous case \cite{Boehm:2020jwd}), potentially applicable to isolated BHs or binary systems, provided there exists an external matter or energy distribution acting as a mediator. In this context, while the dark halo provides a well-motivated astrophysical scenario for SMBHs, it should be regarded as a physical facilitator of the coupling rather than its exclusive fundamental origin. The dark sector model we investigate serves as a concrete realization of this broader universal mechanism.

In this paper, we construct an exact analytical solution for a cosmological coupled BH immersed in an anisotropic dark sector background. By adopting a geometry-driven approach from Ref.\cite{Cadoni:2023lqe, Cadoni:2023lum}, we start from the static seed metric supported by the dark sector. We generalize this static seed to a dynamical cosmological setting. Our model successfully realizes the cosmological coupling, where the BH mass grows with the scale factor. We derive the explicit form of the radius-dependent coupling exponent $k(r)$, which bridges the local BH geometry and the global expansion. Furthermore, we perform a rigorous asymptotic analysis of the horizon structure. Our derivation refines the understanding of the horizon structure within this geometry-driven framework. We demonstrate that for our solution, the apparent horizon remains outside the static boundary, namely $r_{AH}>r_+$, a feature consistent with the conclusion of previous studies that have explored the possibility of apparent horizons extending into the regular region \cite{Cadoni:2024rri}.

The paper is organized as follows. In Sec.II, we briefly review the static BH solution in the dark sector. In Sec.III, we construct the dynamical cosmological solution and derive the coupling factor. In Sec.IV, we analyze the physical properties of the solution, including the matter content, the Misner-Sharp mass evolution, and the apparent horizon. Finally, we discuss our results and conclude in Sec.V. Throughout this paper, we use the geometric units with $G= c = 1$.

\section{Review of the Static Solution}
In this section, we briefly review the static spherically symmetric BH solution immersed in a dark sector background, as originally derived in Ref.\cite{Li:2012zx}. This solution serves as the seed metric and provides the boundary conditions for the dynamical extension constructed in the subsequent sections.

We consider a gravitational system coupled to a dark sector\footnote{One can refer to part B in Sec.III or to Ref.\cite{Liang:2023jrj}, where the reason for calling it the dark sector is explained.} consisting of a PFDM component and a phantom field. The action is given by
\begin{equation}
    S=\int d^{4}x \sqrt{-g} \left( \frac{R}{2\kappa^{2}}+\mathcal{L}_{\rm dark} \right),\label{act} 
\end{equation}
where $\kappa^{2}=8\pi G$, and $\mathcal{L}_{\rm dark}$ is the action of dark sector. We start with the general static spherically symmetric metric ansatz
\begin{equation}
    ds^{2}=-e^{\nu(r)}dt^{2}+e^{\mu(r)} dr^{2}+r^{2}(d\theta^{2}+\sin^{2}\theta d\phi^{2}). \label{ans}
\end{equation}
We assume that the stress-energy tensor of the dark sector is in the form of
\begin{equation}
    T_{\nu}^{\mu }=\mathrm{diag}(-\rho, p_r, p_{\theta}, p_{\phi}) .
\end{equation}
Here, $\rho, p_r, p_{\theta}$ and $p_{\phi}$ represent the energy density, radial pressure, tangential pressure, and poloidal pressure of the total dark sector, respectively.

Under the ansatz \eqref{ans}, the non-vanishing components of the Einstein field equations are explicitly given by
\begin{subequations} 
\begin{align}
e^{-\mu}(\frac{1}{r^{2}}-\frac{\mu'}{r})-\frac{1}{r^{2}}&=-\kappa^2 \rho,\label{tt}\\
e^{-\mu}(\frac{1}{r^{2}}+\frac{\nu'}{r})-\frac{1}{r^{2}}&=\kappa^2 p_r,\label{rr}\\
 \frac{e^{-\mu}}{2}(\nu''+\frac{\nu'^{2}}{2}+\frac{\nu'-\mu'}{r}-\frac{\nu' \mu'}{2})&=\kappa^2 p_{\theta}=\kappa^2 p_{\phi},\label{cc}
\end{align}
\end{subequations}
where the prime $'$ is the derivative with respect to $r$. Comparison of the radial and tangential equations immediately reveals the anisotropic nature of the source ($p_r \neq p_{\theta}$), which is very important for us to construct a cosmological coupled BH solution hereafter. 

Following the line of Ref.\cite{Li:2012zx}, we focus on the condition $\mu(r)=-\nu(r)$ of the ansatz, which can derive the constraint $\rho =-p_{r}$ from the \eqref{tt} and \eqref{rr}. One can find that there is no constraint on energy density $T_{t}^{t}$ and tangential pressure $T_{\theta}^{\theta}$. Generally, this relation can be written as $T^{\theta}_{\theta}=\gamma T^{t}_{t}$, where $\gamma=0,+1,-1$ are relevant to Schwarzschild, Schwarzschild dS/AdS, RN BH, respectively. However, in this paper, we are interested in the anisotropic dark sector fluid background, so we focus on the $\gamma=-\frac{1}{2}$ case as the choice of previous works considering the potential for observation \cite{Xu:2017bpz, Haroon:2018ryd, Konoplya:2019sns, Jusufi:2019nrn, Narzilloev:2020qtd, Rizwan:2018rgs, Stuchlik:2019uvf, Hendi:2020zyw, Cao:2021dcq, Xu:2016ylr, Shaymatov:2020wtj, Gao:2023ltr, Liang:2023jrj, Abdusattar:2023xxs, Liu:2024qso, Sadeghi:2024wib, Rodriguez:2024jzw, Narzilloev:2024fsw, Das:2023ess}. Specifically, logarithmic potential can produce a flat rotational curve, so the $\gamma=-\frac{1}{2}$ case might naturally support the observational fact of a flat rotational curve without fine-tuning the distribution of dark matter. By utilizing the time component \eqref{tt} and radial component \eqref{cc} of the field equation, one can find the relation
\begin{equation}\label{eq1}  e^{\nu}r^{2}\nu''+e^{\nu}r^{2}\nu'^{2}+3e^{\nu}r \nu'+e^{\nu}-1=0.
\end{equation}
Defining $U:=1-e^{\nu}$, one can find that the above differential equation can be represented in a more concise form as
\begin{equation} 
    r^{2}U''+3rU'+U=0,
\end{equation}
and one can find the solution $U=\frac{2M}{r}+\frac{\lambda}{r} \ln \frac{r}{\lambda}$. Finally, we arrive at the static BH solution immersed in a dark sector background as
\begin{align}\label{metric}
ds_{\mathrm{static}}^2&=-f(r)dt^2+\frac{1}{f(r)}dr^2+r^2d\Omega^2,\\
f(r)&=1-\frac{2M}{r}-\frac{\lambda}{r} \ln \frac{r}{\lambda},\nonumber
\end{align}
where $M$ and $\lambda$ are the integration constants corresponding to the BH mass and the density of the dark sector. This solution describes a BH distorted by the surrounding dark sector halo. The parameter $\lambda$ introduces a logarithmic correction to the standard Schwarzschild potential. The event horizon $r_+$ is determined by the largest root of the equation $f(r_+)=0$, and one can also rewrite the metric function \eqref{metric} as 
\begin{equation}
    f(r)=1-\frac{r_{+}}{r}-\frac{\lambda}{r} \ln \frac{r}{r_{+}}.
\end{equation}
We plot the relation of static horizon radius $r_+$ versus density parameter $\lambda$ in FIG.\ref{lam}. One can find that the static horizon does not monotonously depend on the dark sector parameter, and there exists a critical value $\lambda_{crit}=2M/(e-1)\approx1.16M$ (we show the derivation in Appendix.\ref{a}), which originates from the competition of two parts of the potential term $-\frac{2M}{r}-\frac{\lambda}{r} \ln \frac{r}{\lambda}$. One can also find that when $\lambda > 2M$, the event horizon disappears.

\begin{figure}
    \centering
    \includegraphics[width=0.6\linewidth]{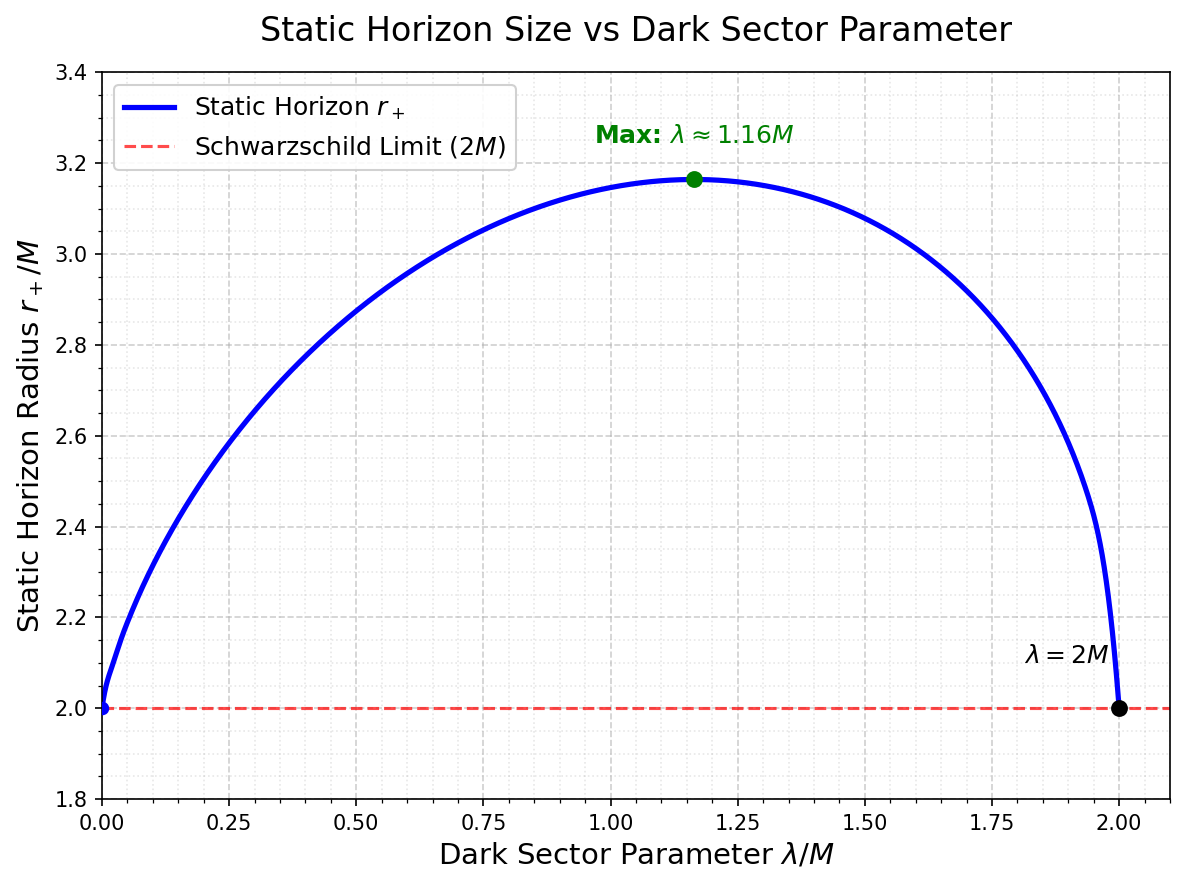}
    \caption{The normalized relation of static horizon radius $r_+$ versus density parameter $\lambda$.} 
    \label{lam}
\end{figure}

Now using the Einstein equations, we can reconstruct the matter distribution profile supporting this static geometry. The energy density is derived from the $(t,t)$ component of the Einstein tensor as
\begin{equation}
\kappa^2\rho(r)=\frac{1}{r^2}\left(1-\frac{d}{dr}[rf(r)]\right).
\end{equation}
Substituting $f(r)$ from \eqref{metric}, we can find
\begin{equation}\label{rho}
\rho(r)=\frac{\lambda}{\kappa^2r^3}.
\end{equation}
Similarly, the radial and tangential pressures are given by
\begin{equation}\label{p}
p_r(r)=-\frac{\lambda}{\kappa^2r^3}, \quad
p_{\theta}(r)=p_{\phi}(r)=\frac{\lambda}{2\kappa^2r^3}.
\end{equation}
These profiles satisfy the weak energy condition provided $\lambda>0$, but violate the strong energy condition due to the phantom-like nature of the background field. The static profiles derived here, $ \rho(r), p_r(r), p_{\theta}(r)$, and $p_{\phi}(r)$ will serve as the \textit{seed} functions for the effective anisotropic fluid in the dynamical embedding framework presented in the next section.

\section{the cosmological coupling solution Construction}
In the previous section, we reviewed the static BH solution immersed in a dark sector background. While the static solution provides valuable insights into the equilibrium of the system, it fails to capture the interactions between the local BH and the evolving universe.

In this section, we generalize the static solution to a dynamical cosmological background. We adopt a geometry-driven framework for embedding compact objects in an expanding universe. By modeling the dark sector as an effective anisotropic fluid and imposing physical boundary conditions, we derive an exact analytical metric that describes the evolution of a cosmological coupling BH immersed in a dark sector background.
\subsection{A generalized cosmological coupling of compact objects framework}
To describe a spherical inhomogeneity embedded in an expanding universe, we employ a generalized time-dependent metric ansatz framework applied in Ref.\cite{Cadoni:2023lqe, Cadoni:2023lum}. We work in conformal time $\eta$ and comoving radial coordinate $r$. The line element is assumed to take the following diagonal form
\begin{equation}
ds^2 = a^2(\eta) \left[ -e^{\alpha(\eta, r)} d\eta^2 + e^{\beta(\eta, r)} dr^2 + r^2 d\Omega^2 \right],
\label{dynamic_ansatz}
\end{equation}
where $a(\eta)$ is the cosmological scale factor with $\alpha(\eta, r)$ and $\beta(\eta, r)$ are metric potentials to be determined. This ansatz has two key advantages, i) Cosmological Limit: at large distances ($r\rightarrow \infty$), if the local potentials vanish ($\alpha, \beta \rightarrow 0$), the metric asymptotically approaches the standard FLRW form. ii) Static Limit: When the scale factor becomes constant ($a(\eta)\rightarrow 1$), the metric can recover the static BH solution like \eqref{ans}.

In order to match the static BH seed, we set $\alpha=\alpha(r)$, which means that $\alpha$ does not explicitly depend on time ($\dot \alpha=0$. Hereafter, the prime ``\,$\prime$\,'' denotes $ \partial/\partial r$, the dot ``$\,\cdot$\,'' denotes $ \partial/\partial \eta$). By computing the Einstein tensor components for the metric \eqref{dynamic_ansatz}, we obtain the following system of coupled differential equations
\begin{subequations} 
\begin{align}
-\frac{1}{r} \dot{\beta} - \mathcal{H} \alpha' &= \kappa^2 T_{\eta r},
\label{field_flux}\\
\frac{1}{a^2} \left[ e^{-\beta} \left( \frac{1}{r^2} - \frac{\beta'}{r} \right) - \frac{1}{r^2} + \mathcal{H}^2 + 2\dot{\mathcal{H}} \right] &= \kappa^2 T_{\eta}^{\eta},
\label{field_rho}\\
\frac{1}{a^2} \left[ e^{-\beta} \left( \frac{1}{r^2} + \frac{\alpha'}{r} \right) - \frac{1}{r^2} + \mathcal{H}^2 + 2\dot{\mathcal{H}} \right] &= \kappa^2 T_r^r,
\label{field_pr}\\
\frac{1}{a^2} \left[ \frac{e^{-\beta}}{2} \left( \alpha'' + (\alpha')^2 - \alpha'\beta' + \frac{\alpha' - \beta'}{r} \right) \right] + \left( \mathcal{H}^2 + \frac{2\ddot{a}}{a} \right) e^{-\alpha}&=\kappa^2 T_{\theta}^{\theta}.
\label{field_pt}
\end{align}
\end{subequations}
where we define the conformal Hubble parameter $\mathcal{H}:=\frac{\dot a}{a}$. Comparing \eqref{field_pr} and \eqref{field_pt}, it is evident that the geometric terms contributing to the radial pressure $T_r^r$ and tangential pressure $T_{\theta}^{\theta}$ are structurally different. The radial component depends only on the first derivatives $\alpha'$, whereas the tangential component involves second derivatives $\alpha ''$. For a general static seed metric 
$f(r)$, the condition of isotropy $T_r^r=T_{\theta}^{\theta}$ would impose an additional differential constraint on the function $f(r)$, which is generally incompatible with the presence of a non-trivial dark sector. Therefore, to construct a consistent dynamical solution that recovers the seed metric in the static limit, the supporting fluid must be anisotropic, i.e. $T_r^r\neq T_{\theta}^{\theta}$. This mathematical requirement of an anisotropic fluid ($p_r\neq p_\theta$) perfectly aligns with the recent understanding that external matter distributions exhibiting such effective anisotropic stresses act as the fundamental mediators of cosmological coupling for general compact objects \cite{Cadoni:2024jxy}. This guarantees that this mechanism is not intrinsically limited to dark matter, but applies to any local system enveloped by suitable energy distributions.

%The equation for $T_{\theta}^{\theta}$ is determined by the angular component $G_{\theta}^{\theta}$, which involves the second derivatives of the metric functions. For brevity, we omit its explicit form here, as it is guaranteed to be satisfied via the conservation law $\nabla_\mu T^{\mu\nu}=0$.

We assume that the total energy-momentum tensor of the system $T^{\mu\nu}$ is diagonal in the comoving frame defined by the metric \eqref{dynamic_ansatz} as $T_{\nu}^{\mu }=\mathrm{diag}(-\rho, p_r, p_{\theta}, p_{\phi})$. This assumption implies that there is no net radial heat flux or momentum flow between the BH and the cosmological background, i.e., $T_{\eta r}=0$. This condition effectively describes a scenario where the BH is comoving with the cosmic expansion, without significant accretion or peculiar velocity (a well-known counter-example is the Sultana-Dyer solution \cite{Sultana:2005tp}).
Imposing the radial no-flux condition $T_{\eta r}=0$ on \eqref{field_flux}, we obtain the differential equation
\begin{equation}
\frac{\partial \beta}{\partial \eta} = -r \alpha' \mathcal{H}.
\label{beta_diff}
\end{equation}
From the above condition $\dot \alpha=0$, we know that $\alpha$ does not explicitly depend on time. Then, we integrate it with respect to time
\begin{equation}
\beta(\eta, r) = -r \alpha'(r) \int \mathcal{H} d\eta + C(r) = -r \alpha'(r) \ln a(\eta) + C(r),
\end{equation}
where the $C(r)$ is an integral function that depends solely on the radial coordinate $r$. Taking the exponential of this result, we derive the general form of the radial metric component
\begin{equation}
e^{-\beta(\eta, r)} = g(r) a(\eta)^{r \alpha'(r)},
\label{beta_solution}
\end{equation}
where $g(r):=e^{-C(r)}$ is a new defining function to absorb the integral, which is finally determined by the spatial boundary conditions. The constraint \eqref{beta_solution} indicates that once the metric potential $\alpha(r)$ is determined, the time evolution of $\beta(\eta,r)$ is completely fixed by the scale factor $a(\eta)$, with no additional degrees of freedom. Now we have a general dynamic ansatz as 
\begin{equation}
ds^2 = a^2(\eta) \left[ -e^{\alpha(\eta, r)} d\eta^2 + \frac{1}{g(r) a(\eta)^{r \alpha'(r)}} dr^2 + r^2 d\Omega^2 \right].
\label{dynamic_ansatz1}
\end{equation}

\subsection{Effective fluid description}
Before proceeding to determine the specific forms of $\alpha(r)$ and $g(r)$, it is necessary to clarify the physical nature of the matter source in this dynamical setting.
 
In the static case discussed in Sec.II, the solution was first derived by a scalar field $V(\Phi)$ with a specific potential coupled to dark matter, which resulted in a specific anisotropic equation of state (one may refer to Ref.\cite{Li:2012zx} for more details). Extending such a specific microscopic scalar field configuration to a dynamic background is mathematically nontrivial and often leads to over-constrained systems if one insists on a rigid form of $V(\Phi)$.

To circumvent these difficulties, we adopt a phenomenological effective fluid approach. Instead of solving the Klein-Gordon equation for the scalar field directly, we treat the entire dark sector (e.g., Phantom field + PFDM) as an effective anisotropic fluid, similarly to what we did in Sec.II. 

In the next part, we will impose two conditions on this fluid to seek consistency of the BH solution, i) Static Correspondence: In the limit of a static universe $a\rightarrow 0$, the fluid stress-energy tensor must strictly recover the density and pressure profiles of the static solution derived in Sec.II. ii) Geometric Scaling: The time evolution of the fluid is determined by the geometric scaling required by the Einstein equations in the expanding background.This procedure ensures consistency because the Einstein tensor $G^{\mu\nu}$ automatically satisfies the Bianchi identities, i.e. $\nabla_\mu G^{\mu\nu}\equiv0$. Therefore, defining the matter source via the geometry as $T_{\mu\nu}=\kappa^{-2}G_{\mu\nu}$ guarantees that the resulting effective fluid satisfies the covariant conservation laws $\nabla_\mu T^{\mu\nu}=0$. This approach allows us to construct an exact solution where the dark sector ``halo" naturally thins with the cosmological expansion.

\subsection{Matching with the static seed}
The general dynamic metric derived in the previous section contains two undetermined spatial functions $\alpha(r)$ and $g(r)$, as well as the cosmological scale factor $a(\eta)$. To construct a dynamic solution, we must fix these functions by imposing rigorous physical boundary conditions. 

We invoke the correspondence principle that in the limit where the cosmological expansion is negligible, our dynamic solution must smoothly reduce to the known static solution derived in Sec.II. This ensures that the local physics features of the BH (e.g., its mass parameter $M$ and dark sector parameter $\lambda$) are correctly embedded into the dynamic framework. Mathematically, the static limit corresponds to fixing the scale factor to unity as
\begin{equation}
a(\eta) \to 1, \quad \dot{a} \to 0, \quad \ddot{a} \to 0.
\end{equation}
In this limit, the conformal time $\eta$ coincides with the cosmic time $t$, i.e. $dt=ad\eta\rightarrow d\eta$. Now, let us recall the general dynamic ansatz \eqref{dynamic_ansatz1} under this limit as
\begin{equation}
ds^2 \xrightarrow{a \to 1} -e^{\alpha(r)} dt^2 + \frac{1}{g(r)} dr^2 + r^2 d\Omega^2.
\label{limit_metric}
\end{equation}
It is crucial to emphasize that $\alpha(r)$ and $g(r)$ emerge from the dynamic field equations \eqref{field_flux}-\eqref{field_pt} as integration functions. From a purely mathematical standpoint, they are arbitrary. However, the physical legitimacy of the solution is guaranteed by the boundary condition of the cosmological expansion velocity. By requiring that the spacetime geometry must continuously deform into the static BH solution \eqref{metric} when the expansion rate vanishes $\mathcal{H} \rightarrow 0$, we uniquely fix these integration functions. This physical continuity requirement ensures that our dynamic metric is not an arbitrary choice, but the \textit{unique} time-dependent generalization of such BHs within the cosmological coupling of compact objects framework.

Then, by comparing the metric components of \eqref{metric} and \eqref{dynamic_ansatz1}, we can uniquely identify the unknown potentials. Matching the gravitational potential to the static redshift factor gives
\begin{equation}
e^{\alpha(r)} = f(r) \quad \implies \quad \alpha(r) = \ln f(r).
\end{equation}
This identification physically implies that the gravitational potential $\alpha$ in the comoving frame is determined entirely by the static mass distribution of the BH and its halo. Simultaneously, matching the radial components in the static limit gives
\begin{equation}
\frac{1}{g(r)} = \frac{1}{f(r)} \quad \implies \quad g(r) = f(r),
\end{equation}
which implies that the function $g(r)$, which appears as an integration constant in the radial no-flux constraint, represents the intrinsic spatial curvature of the seed BH solution.

With $\alpha(r)$ and $g(r)$ determined, the full dynamic behavior of the radial metric component $e^{\beta(\eta,r)}$ is now fixed. Substituting these results back into the general constraint derived in \eqref{beta_solution}, we will find
\begin{equation}
e^{\beta(\eta, r)} = \frac{a(\eta)^{-r \frac{d}{dr}(\ln f(r))}}{f(r)}.
\end{equation}
Here, a critical new quantity emerges naturally from the derivation. We define the radius-dependent coupling exponent as
\begin{equation}
k(r) \equiv r \alpha'(r) = \frac{r f'(r)}{f(r)}.
\end{equation}
We call it the cosmological coupling factor, which was first described in Ref.\cite{Cadoni:2023lqe, Cadoni:2023lum}. This procedure is not merely a mathematical substitution; it establishes the interaction rule between the local object and the global universe. The function $f(r)$ describes the seed shape of the BH solution. Meanwhile, the exponent $k(r)$ describes how strongly this shape is stretched or compressed by the cosmological expansion $a(\eta)$.

By explicitly calculating $f'(r)$ for the metric \eqref{metric}, we obtain the exact analytical form of this coupling as
\begin{equation}\label{kr}
k(r) = \frac{2M + \lambda (\ln \frac{r}{\lambda} - 1)}{r - 2M - \lambda \ln \frac{r}{\lambda}}.
\end{equation}
This specific form of $k(r)$ dictates that the coupling is non-uniform: it vanishes at spatial infinity (i.e. $k\rightarrow 0$ as $r\rightarrow \infty$), recovering the FLRW limit, but becomes significant near the BH horizon, leading to non-trivial dynamical effects. When near the static event horizon, we can Taylor-expand the seed metric function as $f(r)\approx f'(r_+)(r-r_+)$, and the coupling exponent can be expressed as $k(r)\approx \frac{r_+}{r-r_+}$. One can find that the divergent nature of the coupling $k(r)$ is a universal geometric feature of the singular horizon, independent of the halo density $\lambda$. To understand the effect of the density parameter $\lambda$ on the coupling exponent $k(r)$, we plot FIG.\ref{k}. For an observer at a fixed coordinate radius $r=3M$, increasing $\lambda$ dramatically enhances the cosmological coupling. Since the maximum extent of the static horizon ($r_+\approx3.16M$ at $\lambda\approx1.16M$) exceeds the observer's position $3M$. Consequently, the coupling exponent $k(r)$ diverges monotonically. This also confirms the divergent behavior of the coupling $k(r)$ near the static event horizon. However, in the $r=4M$ case, the observer is located safely outside the maximum reach of the horizon. As $\lambda$ increases towards the critical value $1.16M$, the horizon expands and moves closer to the observer, causing $k$ to rise. However, once $\lambda$ exceeds the critical value, the horizon begins to shrink and recede away from the observer (one can also refer to FIG.\ref{lam} to associate such behavior). Such behaviors brought by $k(r)$ imply the complicated nonlinear coupling relation between the halo matter and the cosmic expansion.

Here, while the local behavior of $k(r)$ presented in FIG.\ref{k} is mathematically informative for understanding the geometric stretching of the seed metric, it is crucial to clarify its physical interpretation. From a physical standpoint, a local observer does not experience this cosmological coupling directly at any given instant; locally, the geometry appears equivalent to the static seed metric. The function $k(r)$ acts purely as a geometric exponent dictating the strength of the coupling at a specific spatial scale. The actual physical manifestation of cosmological coupling can only be inferred by comparing local gravitational systems across different cosmological epochs, i.e., at different values of the scale factor $a$. Therefore, the physically relevant observable is not $k(r)$ itself, but the evolutionary factor $a^{k(r)}$ evaluated at a fixed spatial scale characteristic of the local system. This factor is what finally drives the mass growth of the coupled object as the universe expands.
\begin{figure}
    \centering
    \includegraphics[width=0.5\linewidth]{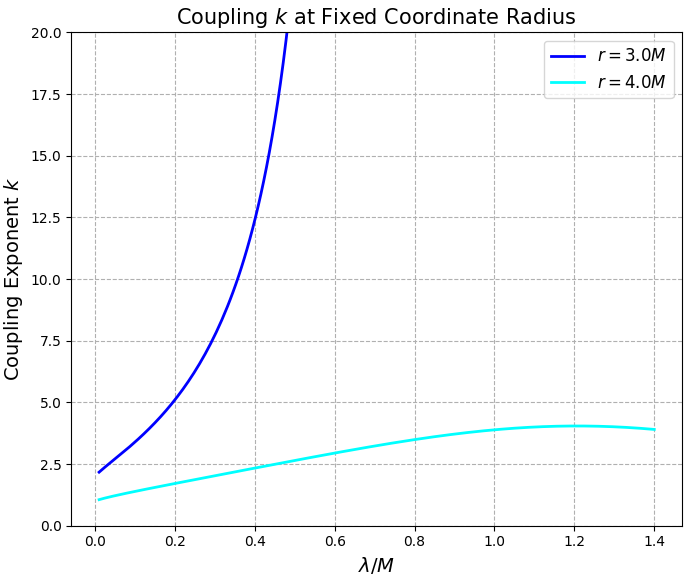}
    \caption{The density parameter $\lambda$ versus the coupling exponent $k(r)$ at a fixed coordinate radius. The lines do not cross at the starting point because in the Schwarzschild limit $\lambda\rightarrow 0$, $k(r)_{\lambda\to0}\sim\frac{2M}{r-2M}$, i.e. $k\sim2$ for $r=3M$ and $k\sim1$ for $r=4M$.} 
    \label{k}
\end{figure}

We finally arrive at a cosmological coupled BH solution written as
\begin{equation}
ds^2_{\text{dyn}} = a^2(\eta) \left[ -f(r) d\eta^2 + \frac{1}{f(r) a(\eta)^{k(r)}} dr^2 + r^2 d\Omega^2 \right].
\label{dynamic_ansatz2}
\end{equation}
where $f(r)$ and $k(r)$ are described in \eqref{metric} and \eqref{kr}, respectively.

\section{Physical Properties}
\subsection{Matter content}
In this part, we analyze the physical properties of the matter source required to support the cosmological coupled geometry. We treat the dark sector as a single effective anisotropic fluid.

We begin by computing the Einstein tensor $G^{\mu}_{\nu} $ for our derived dynamic metric \eqref{field_flux}-\eqref{field_pt}, and the time component of the field equation is now given as
\begin{equation}\label{exact_rho}
\kappa^2 \rho(\eta, r) = \frac{\mathcal{H}^2}{a^2 f(r)} \left( 3 - k(r) \right) + \frac{1}{a^2 r^2} \left[ 1 - a(\eta)^{k(r)} \Big( (rf(r))' + f(r) r k'(r) \ln a(\eta) \Big) \right],
\end{equation}
and one can find that the cosmological background part is modulated by $(1-3k)/f$. The linear dependence on $k(r)$ indicates a coupling between the BH and the local expansion rate. In the limit of $M,\lambda\rightarrow 0$, i.e., $f\rightarrow1,k\rightarrow0$, which means the Minkowski limit of the seed metric and no coupling, the field equation will recover the pure FLRW background. The local density profile contains the term $k'(r) \ln a(\eta)$, which represents a time-dependent distortion of the halo shape. In the static limit $a\rightarrow1$, $\mathcal{H}\rightarrow0$, the expression reduces to $\kappa^2 \rho=\frac{1}{r^2}(1-(rf)')$, and one can easily find that $(rf)'=1-\lambda/r$, which matches the static density \eqref{rho}.

The radial pressure $p_r$ corresponds to the $(r,r)$ component, and the field equation can be expressed as 
\begin{equation}
\kappa^2 p_r(\eta, r) = -\frac{1}{a^2 f(r)} \left( 2\dot{\mathcal{H}} + \mathcal{H}^2 \right)+ \frac{1}{a^2 r^2} \left[ a(\eta)^{k(r)} (rf(r))' - 1 \right].
\label{exact_pr}
\end{equation}
To verify the validity of this expression, we take the static limit $a\rightarrow1$, and the cosmological term vanishes, the radial pressure reduces to $\kappa^2 p_{r} = \frac{1}{r^2} \left[ (rf)' - 1 \right]$, which coincides with the static pressure \eqref{p}. The tangential pressure $p_{\theta}$ reveals the complexity of the anisotropic stress, and it can be written as 
\begin{equation}
\begin{aligned}
\kappa^2 p_\theta(\eta, r) &= -\frac{1}{a^2 f(r)} \left[ (2-k(r))\dot{\mathcal{H}} + \left(1 - k(r) + \frac{1}{2}k(r)^2\right)\mathcal{H}^2 \right] \\
& \quad + \frac{f(r) a(\eta)^{k(r)}}{2 a^2} \left[ \frac{f''(r)}{f(r)} + \frac{2f'(r)^2}{f(r)^2} + \frac{2f'(r)}{rf(r)} + k'(r) \ln a(\eta) \left( \frac{1}{r} + \frac{f'(r)}{f(r)} \right) \right].
\end{aligned}
\label{exact_pt}
\end{equation}
Compared with \eqref{exact_pr}, the coefficient of $\mathcal{H}^2$ in $p_r$ is $1$, whereas in $p_\theta$ it is $1-k+\frac{1}{2}k^2$, which shows strong background anisotropy, confirming that $p_r\neq p_\theta$ is a necessary condition for this solution.

To check the physical consistency, we also examine the local equation of state. Summing the density and radial pressure
\begin{equation}
 \rho(\eta,r) + p_{r}(\eta,r) = -\frac{f(r) a(\eta)^{k(r)}}{\kappa^2 a^2 r} \cdot k'(r) \ln a(\eta).
\end{equation}
This confirms that while the static seed satisfies $\rho=-p_r$, the dynamical embedding introduces a deviation proportional to $k' \ln a(\eta)$. This deviation represents the dynamical stress required to maintain the BH geometry against the cosmological expansion shear. The solution is physically viable as this stress vanishes at spatial infinity $r\rightarrow\infty$ and remains finite outside the horizon, and also clearly satisfies the static limit.

\subsection{Misner-Sharp Mass}
For BHs in asymptotically flat spacetimes with conserved and finite mass, the energy is well defined by the ADM mass in terms of a timelike integral at spatial infinity on a 2D sphere. However, characterizing the evolution of dynamical spacetimes is a complex problem. In our case, the spacetime is an asymptotically FLRW solution at large distances, and the ADM mass cannot be defined in non-asymptotically flat spacetimes. To quantify the energy content of the dynamical BH and investigate the interaction between the local object and the cosmological background, it is proper to apply the Misner-Sharp mass to define a quasi-local energy \cite{Misner:1964je}, which naturally defines the total energy contained within a sphere of a given size in a spherically-symmetric spacetime \cite{Hayward:1994bu}.

The Misner-Sharp mass is defined covariantly by the gradient of the physical radius $R_L:=a(\eta) r$ as
\begin{equation}
M_{\rm MS}(\eta, r) = \frac{R_L}{2} \left( 1 - g^{\mu\nu} \partial_\mu R_L \partial_\nu R_L \right),
\label{MS_definition}
\end{equation}
where $g^{\mu\nu}$ is the inverse metric. For the metric \eqref{dynamic_ansatz2}, we have $g^{\eta\eta} = -{(a^2 f(r))}^{-1}, g^{rr} = {f(r) a(\eta)^{k(r)-2}}$. The partial derivatives of the chosen radius are $\partial_\eta R_L = \dot{a} r, \partial_r R_L = a$. Substituting these components into the gradient part, we have
\begin{align}
g^{\mu\nu} \partial_\mu R_L \partial_\nu R_L &= g^{\eta\eta} (\partial_\eta R_L)^2 + g^{rr} (\partial_r R_L)^2 \nonumber \\
&= -\frac{1}{a^2 f(r)} (\dot{a} r)^2 + \frac{f(r) a(\eta)^{k(r)}}{a^2} (a)^2 \nonumber \\
&= -\frac{\mathcal{H}^2 r^2}{f(r)} + f(r) a(\eta)^{k(r)},
\label{gradient_R}
\end{align}
Inserting \eqref{gradient_R} into the definition \eqref{MS_definition} and rearranging the terms, we finally arrive at the exact analytical expression for the MS mass
\begin{equation}
 M_{\rm MS}(\eta, r) = \frac{\mathcal{H}^2 a(\eta) r^3}{2 f(r)}+\frac{a(\eta) r}{2} \left[ 1 - f(r) a(\eta)^{k(r)} \right]  . 
\label{MS_final}
\end{equation}

Notably, one may find that the expression for $M_{\rm MS}$ can be physically interpreted by decomposing it into two parts as a cosmological background contribution plus a coupled BH contribution. The first term in \eqref{MS_final} represents the energy of the cosmic fluid enclosed within the sphere
\begin{equation}
M_{\text{MS,cos}} = \frac{\mathcal{H}^2 a r^3}{2 f(r)} = \frac{4\pi}{3} R_L^3 \left( \frac{3 H^2}{8\pi} \right) \frac{1}{f(r)},
\end{equation}
where $H=\mathcal{H}/a(\eta)$ is the physical Hubble parameter. In the limit of the seed metric $f(r)\rightarrow1$, this part reduces to $\sim \frac{4 \pi}{3} \rho_c R_L^3$, which is simply the energy of a sphere filled with the background density $\rho_c\sim\frac{3 H^2}{8\pi}$ in a FLRW universe. When near the BH, the factor ${f(r)}^{-1}$ will significantly enhance this mass. This reflects the distortion of the background by the central object. 

The second term in \eqref{MS_final} describes the intrinsic mass within the halo modified by the cosmological expansion
\begin{equation}
M_{\text{MS,halo}} = \frac{a r}{2} \left[ 1 - f(r) a(\eta)^{k(r)} \right].
\end{equation}
To understand the nature of the coupling, one can apply the static MS mass within the halo $\mathcal{M}_{ms}(r)=M+\frac{\lambda}{2}\ln(r/\lambda)$ \cite{Liang:2023jrj} from the metric \eqref{metric}, then expanding the terms
\begin{align}
M_{\text{MS,halo}} &= \frac{a r}{2} \left[ 1 - \left( 1 - \frac{2\mathcal{M}_{ms}(r)}{r} \right) a^{k} \right] \nonumber \\
&= \frac{a r}{2} (1 - a^k) + \mathcal{M}_{ms}(r) a^{1+k} .
\end{align}
One may identify that the first term originates from the local curvature, and the second term $\mathcal{M}_{ms}(r) \cdot a(\eta)^{1 + k(r)}$ directly shows the coupling with the universe expanding. Our model exhibits strong non-linear coupling, implying that the additional dark halo renders the BH mass more sensitive to cosmic expansion, involving more complex radial dependence due to the logarithmic terms in the dark sector potential.

\subsection{Apparent Horizon}
In dynamical spacetimes, the event horizon is not properly defined, but only the apparent one has physical meaning \cite{Visser:2014zqa}. One can also refer to a recent study \cite{Faraoni:2024ghi}, which demonstrates that forcing an exactly static and spherically symmetric BH event horizon embedded in a time-dependent geometry is physically untenable, resulting in an additional naked singularity. This implies that it would be appropriate to choose the apparent horizon as the physical boundary for the cosmological BHs. The apparent horizon is defined as the marginal trapped surface where the expansion of the outgoing null geodesics vanishes. In spherically symmetric spacetimes, this condition is geometrically equivalent to the surface where the gradient of the physical radius $R_L$ becomes null
\begin{equation}
g^{\mu\nu} \nabla_\mu R_L \nabla_\nu R_L = 0.
\label{AH_def}
\end{equation}

In the previous part, while deriving the Misner-Sharp mass, we have calculated the norm of the gradient vector $\nabla R_L$, seeing \eqref{gradient_R}. The location of the apparent horizon, denoted by the comoving apparent radius $r_{\rm AH}$, is determined by setting this scalar invariant to zero as
\begin{equation}
\left( \frac{f(r_{\rm AH})}{r_{\rm AH}} \right)^2 a(t)^{k(r_{\rm AH})-2} = H(t)^2 ,
\label{AH_final}
\end{equation}
\begin{equation}
\left( \frac{1-\frac{2M}{r}-\frac{\lambda}{r_{\rm AH}} \log \frac{r_{\rm AH}}{\lambda}}{r_{\rm AH}} \right)^2 a(t)^{\frac{2M + \lambda (\ln \frac{r_{\rm AH}}{\lambda} - 1)}{r_{\rm AH} - 2M - \lambda \ln \frac{r_{\rm AH}}{\lambda}}-2} = H(t)^2 .
\end{equation}
Considering the complicated form of the expression $f(r)$ and $k(r)$, we cannot explicitly express the analytical form of the apparent horizon. Here, we examine the asymptotic behavior again. For the static limit $a\rightarrow1, H\rightarrow0$, the RHS of \eqref{AH_final} becomes $0$, and the LHS becomes $f(r_{\rm AH})/r_{\rm AH}=0$, resulting in $r_{\rm AH}=r_+$, which means the apparent horizon coincides with the event horizon. In the FLRW limit $f\rightarrow1, k\rightarrow0$, there exists the relation $1/(a(t)r)=H(t)$, i.e., $R_L=1/H$ which is consistent with the Hubble horizon.

\begin{figure}
    \centering
    \includegraphics[width=1\linewidth]{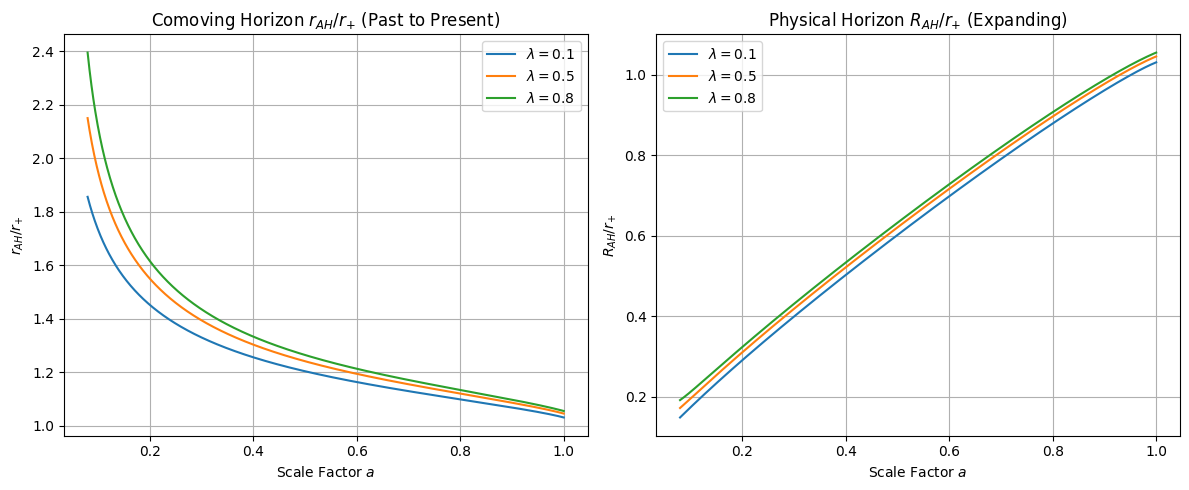}
    \caption{Numerical evolution of the apparent horizon. Left Panel: The normalized comoving radius versus static event horizon $r_{\rm AH}/r_+$ as a function of the scale factor. Right Panel: The normalized physical areal radius versus static event horizon $R_{\rm AH}/r_+$, showing the global increase of BHs. Note that at the present epoch $a=1$, the normalized horizon radius slightly deviates from the point $(1, 1)$. This is because we assume a non-static universe with initial expansion rate $H\neq0$. In the strict static limit $H\rightarrow0$, the ratio asymptotically converges to exactly $1$.} 
    \label{horizon}
\end{figure}
For further investigation, we numerically plot the evolution of the apparent horizon in an expanding universe in FIG.\ref{horizon}. We assume a standard $\Lambda$CDM background characterized by the Hubble parameter $H(a)\propto \sqrt{\Omega_m a^{-3}+\Omega_\Lambda}$ where $\Omega_m = 0.3, \Omega_\Lambda = 0.7$. Here, we take the $\lambda<M$ case into account. This is because for standard astrophysical (stellar or supermassive) BHs, the surrounding dark halo density, even accounting for a spike near the horizon, is typically orders of magnitude lower than the BH's characteristic density. We can find that the comoving apparent horizon $r_{\rm AH}$ is always located outside the static event horizon $r_+$. Such behavior is a direct consequence of the divergent coupling exponent $k(r)$ at the static horizon, making $(\nabla R_L)^2<0$ (trapped) always within the static horizon $r_+$. Despite the comoving contraction, the physical areal radius increases monotonically (see the right panel). This confirms that the global cosmic expansion dominates over the local contraction, consistent with the observed growth of SMBHs over cosmological timescales. A comparison of the curves reveals a direct relationship between the dark sector density $\lambda$ and the growth rate. The BH with a dense halo $\lambda=0.8$ grows faster than the one with a thin one $\lambda=0.1$. This suggests that the anisotropic fluid halo amplifies the response of the horizon geometry to the stretching force of the Hubble flow.

It is worth noting that our result $r_{\rm AH}>r_+$ confirms the conclusion presented in Ref.\cite{Cadoni:2024rri}, which demonstrates that the apparent horizon is always located outside the static event horizon for a cosmological coupled BH. This behavior can be confirmed by examination of the asymptotic behavior of the metric functions. The location of the apparent horizon is determined by the condition \eqref{AH_def} as $|f(r)| = \mathcal{H} r \, a^{-k(r)/2}$ \footnote{Note, one should distinguish between the cosmological apparent horizon and the BH apparent horizon. For $r\gg r_+$, the above equation reduces to $1=\mathcal{H} r$ and is consistent with a cosmological horizon with physical radius $R_{\rm L, cos}=1/H(t) $. Here, we only focus on the BH apparent horizon in the present work. } (here we convert \eqref{AH_final} to conformal Hubble parameter for a concise form), and for the past universe where $a < 1$. Consider the behavior immediately outside the static horizon ($r \to r_+^+$). In this region, $f(r) > 0, \rightarrow0$ and $f'(r) > 0$ (Whether the Fan \& Wang and Hayward metric used in Ref.\cite{Cadoni:2024rri} or the static seed metric \eqref{metric} presented in our paper, the metric function is monotonically increasing just outside the event horizon), which implies $k(r)=rf'/f \to +\infty$. Consequently, the term on the right-hand side of the above relation, $a^{-k(r)/2}$, diverges exponentially to positive infinity as
\begin{equation}
    a^{-k(r)/2} \sim \exp\left( -\frac{C}{r - r_+} \right) \quad \text{when} \quad r \to r_+^+ \,,
\end{equation}
where $C < 0$ is a constant determined by the event horizon and the scale factor ($C \propto r_+ f'(r_+) \ln a$). Since the left-hand side $f(r)$ approaches zero while the effective Hubble flow term diverges, the intermediate value theorem guarantees that the curves must intersect at some radius $r_{\rm AH}>r_+$. Readers can feasibly verify this relationship by numerically plotting the intersection of $|f(r)|$ and the effective Hubble flow term $\mathcal{H} r a^{-k(r)/2}$. 

It is crucial to emphasize that this result does not conflict with the conclusion that the event horizon of an expanding BH needs to be outside the apparent horizon. Physically, the BHs with expanding mass featured an apparent horizon inside the event horizon, as explicitly seen in the Vaidya case \cite{Booth:2010eu}. It is a consequence of the teleological nature of the event horizon and is guaranteed by the null energy condition in classical dynamics \cite{Hayward:1993wb}, which leads to the local trapping surface (apparent horizon) being contained within the global causal boundary (event horizon or its static counterpart). However, in our case, the static event horizon radius $r=r_+$ is not a true event horizon for a dynamical BH, but connects with an additional singularity, which we will note in the next part. Overall, as concluded in Refs.\cite{Visser:2014zqa, Faraoni:2024ghi}, the apparent horizon is more suitable than the event horizon as a physical boundary in the cosmological context.

\subsection{Singularity}
To confirm the asymptotic behavior of curvature structure and the nature of the singularities in the dynamical spacetime, we calculate the Kretschmann scalar, defined as
\begin{equation}
\mathcal{K} := R_{\mu\nu\rho\sigma}R^{\mu\nu\rho\sigma}.
\end{equation}
By utilizing the metric \eqref{dynamic_ansatz1} and the definition of the Kretschmann scalar, we get a complicated expression using the xCoba in \textit{Mathematica}, written as
\begin{equation}\label{ks}
\begin{aligned}
\mathcal{K} &= \frac{1}{4 a^8 f^2 r^4} \Bigg\{ 
    16 a^4 f^2 - 32 a^{4+k} f^3 
    + \left(k^4 - 32 k + 96\right) r^4 \dot{a}^4 \\
    &\quad - 4 a \left(k^3 - 2 k^2 - 4 k + 24\right) r^4 \dot{a}^2 \ddot{a} 
    + 4 a^2 \left[ 8 f r^2 \dot{a}^2 + \left(k^2 - 4 k + 12\right) r^4 \ddot{a}^2 \right] \\
    &\quad + 4 a^{3+k} f r^3 \ddot{a} \left[ f' \left( -8 + (k-2) r k' \ln a \right) + 2 (k-2) r f'' \right] \\
    &\quad + a^{4+2 k} f^2 \Bigg[ 
        16 f r^2 f' k' \ln a 
        + 8 f^2 \left( 2 + r^2 (k')^2 (\ln a)^2 \right) \\
        &\qquad + r^2 (f')^2 \left( 16 + r^2 (k')^2 (\ln a)^2 \right) 
        + 4 r^4 f' k' f'' \ln a + 4 r^4 (f'')^2 
    \Bigg] \\
    &\quad - 2 a^{2+k} r^2 \dot{a}^2 \Bigg[ 
        8 r^2 (f')^2 + 8 f^2 \left( k^2 + 2 + (2 - k) r k' \ln a \right) \\
        &\qquad + f r \bigg( -24 k f' - 4 r \left( f' k' \ln a + 2 f'' \right) 
        + k^2 r \left( f' k' \ln a + 2 f'' \right) \bigg) 
    \Bigg] 
\Bigg\}
\end{aligned}
\end{equation}

Firstly, we note the central singularity at $r\rightarrow0$. In this limit, the static seed metric function \eqref{metric} dominates as $f(r)\approx -\frac{\lambda}{r} \ln(\frac{r}{\lambda})$, $f'(r)\approx \frac{\lambda}{r^2} \ln(\frac{r}{\lambda})$ and $f''(r)\approx -\frac{2\lambda}{r^3} \ln(\frac{r}{\lambda})$, respectively. With the coupling exponent behaving as $k(r)\rightarrow-1$, keeping the leading term, the Kretschmann scalar is 
\begin{equation}
\mathcal{K} \sim \frac{12 \lambda^2 (\ln(\frac{r}{\lambda}))^2}{a(\eta)^6 r^6}.
\end{equation}
This confirms that the central singularity at $r=0$ persists, consistent with the singular nature of the seed metric. The singularity is not removed by the cosmological coupling, which is very different from the regular BH cases \cite{Cadoni:2023lqe, Cadoni:2023lum}.

Now, shift our focus to another special position $r=r_+$, connecting with the static seed event horizon, which is not a curvature singularity in the static case. In the limit $r\rightarrow r_+$, the seed metric function $f(r)\rightarrow0$, with $k(r)\rightarrow \infty$. Checking the expression of the Kretschmann scalar \eqref{ks}, we identify two types of divergent contributions. The first is polynomial divergence. Consider the term proportional to $\dot a^4$ inside the curly brackets, $\mathcal{K}_{\text{poly}} \sim \frac{1}{f^2} \cdot k^4 \dot{a}^4 \sim \frac{1}{f^2} \cdot \left(\frac{1}{f}\right)^4 = \frac{1}{f^6}$, which indicates a strong polynomial curvature singularity. The second is exponential divergence. Consider the terms involving the factor $a^{4+2k}$. For an expanding universe $a>1$ and positive coupling near the horizon $k\rightarrow\infty$, this factor introduces an exponential divergence as $\mathcal{K}_{\text{exp}} \sim \frac{a^{2k}}{f^2} \cdot \left[ (f')^2 (k')^2 (\ln a)^2 \right]$. Both confirm that the Kretschmann scalar diverges at $r=r_+$,
\begin{equation}
\mathcal{K}|_{r \to r_+} \to \infty.
\end{equation}
This divergence indicates that the static event horizon $r=r_+$ becomes a curvature singularity in the dynamical spacetime. Physically, this corresponds to a divergence in the effective energy density and pressure at the horizon\footnote{Unlike the standard McVittie solution where the energy density remains finite (spatially homogeneous) and only pressure diverges at the horizon, our solution exhibits divergences in both energy density and pressure. This is due to the non-vacuum nature of the static seed (dark Sector halo) and the anisotropic geometric coupling, which prevents the cancellation of divergent terms in the $G_{00}$ component typically seen in the McVittie case. } (as seen in \eqref{exact_pr}), which inevitably violates the weak energy condition in this limit. This is a known feature of the no-flux condition $T_{\eta r}=0$, which is very similar to the McVittie-type solutions \cite{Faraoni:2012gz}. It implies that the surface $r=r_+$ acts as a rigid, impenetrable boundary for the cosmic fluid, where the energy density and pressure required to maintain the static geometry against the cosmic expansion become infinite.

\section{Conclusion and Discussion}
In this paper, we constructed an exact analytical solution for a cosmological coupled BH immersed in an anisotropic dark sector background. By generalizing a static seed metric, we derived a dynamical solution where the BH mass co-evolves with the cosmic expansion, scaling as $M_{\rm MS}\sim a(\eta)^{k(r)}$, that physically realizes the cosmological coupling phenomenon highlighted by observations of SMBH growth \cite{Farrah:2023opk}.

We show that the cosmological coupling is not an arbitrary parameter but rather is geometrically determined by the profile of the dark sector halo. The logarithmic terms in the seed metric lead to a non-uniform coupling $k(r)$ that becomes dominant in the strong-field regime. This suggests that the above anomalous mass growth observed might be physically interpreted as the response of the surrounding anisotropic dark sector fluid to the Hubble flow, rather than a modification of the BH's internal equation of state \cite{Croker:2021duf, Croker:2019mup}. That is another possible way to match the observation, considering the ubiquity of the dark halos surrounding SMBHs. We stress that restricting the cosmological coupling solely to dark matter halos is not a fundamental limitation of this framework. As recently pointed out in \cite{Cadoni:2024jxy}, cosmological coupling is expected to be a universal property of local gravitational systems. The essential ingredient for such a coupling is the presence of an external matter or energy distribution with anisotropic pressure. Therefore, while the dark sector model provides a potential realization for SMBHs, similar coupling effects are fundamentally feasible for other compact systems, such as isolated or binary BHs, provided the necessary effective anisotropic stress is present in their vicinity.

We also find that the apparent horizon remains strictly confined outside the static boundary, which is consistent with the conclusion of Ref.\cite{Cadoni:2024rri}. Specifically, the static boundary at $r=r_+$ manifests as a curvature singularity with divergent energy density. As often seen in McVittie-type solutions, this singularity should be interpreted physically as a rigid boundary layer where infinite tension is required to counteract the shear of cosmic expansion, preserving the exterior geometry as a viable and stable model for cosmological coupling \cite{Faraoni:2012gz}. But meanwhile, the apparent horizon effectively screens the interior singularity, thereby enabling the thermodynamics on the apparent horizon to be defined.

Finally, it is worth noting that a new approach has recently been proposed by Cadoni et al.\cite{Cadoni:2026ejk} to regularize such singularities by accounting for the backreaction of the local geometry on the cosmological dynamics. Extending our future work to incorporate these backreaction effects is a promising direction.

\appendix
\section{The maximum $\lambda$ in static solution}\label{a}
For the static horizon, we have
\begin{equation}
F(r,\lambda)=1-\frac{2M}{r_+}-\frac{\lambda}{r_+} \log \frac{r_+}{\lambda}.
\end{equation}
We want to know how $r_+$ varies with $\lambda$, that is
\begin{equation}
\frac{d r_+}{d \lambda}=- \frac{\partial F/ \partial \lambda}{\partial F/ \partial r_+},
\end{equation}
with
\begin{equation}
\begin{aligned}
\frac{\partial F}{\partial\lambda}&=-\frac{1}{r_+}\ln\left(\frac{r_+}{\lambda}\right)-\frac{\lambda}{r_+}\cdot\frac{1}{r_+/\lambda}\cdot\left(-\frac{r_+}{\lambda^2}\right)\\
&=\frac{1}{r_+}\left[1-\ln\left(\frac{r_+}{\lambda}\right)\right].
\end{aligned}
\end{equation}
Regarding the inflection points, we have $\frac{d r_+}{d \lambda}=0$ as $1-\ln\left(\frac{r_+}{\lambda}\right)=0$, i.e., $r_+=e\cdot\lambda$. Substituting it back into the horizon equation, we have
\begin{equation}
\begin{aligned}
1-\frac{1}{e}&=\frac{2M}{e\lambda},
\end{aligned}
\end{equation}
and we get the $\lambda_{crit}=\frac{2M}{e-1}$.

\begin{acknowledgments}
This work is supported by the National Natural Science Foundation of China (NSFC) under Grant No.12175105.
\end{acknowledgments}

%%%%%%%%%%%%%%%%%%%%%%%%%%%%%%%%%%%%%%%%%%%%%%%%%%%%%%%%%%%%%
\bibliography{bbb}

\end{document}